# A New Approach to Modeling and Analyzing Security of Networked Systems

Gaofeng Da
Institute for Cyber Security
UT San Antonio
dagfvc@gmail.com

Maochao Xu
Dept. of Mathematics
Illinois State University
mxu2@ilstu.edu

Shouhuai Xu
Dept. of Computer Science
UT San Antonio
shxu@cs.utsa.edu

## ABSTRACT

Modeling and analyzing security of networked systems is an important problem in the emerging Science of Security and has been under active investigation. In this paper, we propose a new approach towards tackling the problem. Our approach is inspired by the *shock model* and *random environment* techniques in the Theory of Reliability, while accommodating security ingredients. To the best of our knowledge, our model is the first that can accommodate a certain degree of *adaptiveness of attacks*, which substantially weakens the often-made independence and exponential attack inter-arrival time assumptions. The approach leads to a stochastic process model with two security metrics, and we attain some analytic results in terms of the security metrics.

## Categories and Subject Descriptors

D.4.6 [**Security and Protection**]

## 1. INTRODUCTION

The long outstanding problem of modeling and analyzing security of networked cyber systems from a whole-system perspective is yet to be tackled satisfactorily. One issue that hinders substantial progress is the difficulty encountered in modeling the adaptiveness of attacks. Indeed, existing models often assumed that the relevant random variables are independent of each other (except [26]), and the attack inter-arrival time follows the exponential distribution. While these assumptions can lead to elegant results, it is important to pursue models that can accommodate weaker assumptions.

In this paper, we propose a new approach to modeling and analyzing security of networked systems. Our model is inspired by the *shock model* and *random environment* techniques in the Theory of Reliability, while accommodating security ingredients. The shock model technique was originally used to describe the phenomenon that systems (or components) may or may not fail due to "shocks" — depending on magnitudes of the shocks [9]. This inspires us to model attack and defense capabilities in the same spirit, namely explicitly considering attack power and defense capabilities. The random environment technique was originally used for describing the external environment that has an impact on the performance of systems, and for explaining the dependence between systems that operate in the same environment [23, 16]. The random environment technique is appealing because it can accommodate a certain degree of *adaptiveness*, which can be seen as the *dynamic* dependence between attacks (rather than the *static* dependence recently investigated in [26]).

### 1.1 Our Contributions

Our contributions are in two-fold. First, we propose a new approach to modeling and analyzing security of networked systems. The approach allows us to accommodate a certain degree of adaptiveness of attacks, namely the dynamic dependence between the relevant random variables. In contrast to a straightforward stochastic model that encounters the state-space-explosion problem, we obtain a $n$-dimensional stochastic model that leads to approximation results or bounds (which are numerically confirmed), where $n$ is the size of the networked system (i.e., number of nodes).

Second, the approach leads to two security metrics, called *time-to-compromise* and *steady-state compromise probability*. The former captures the random time it takes for a secure (but vulnerable) node/computer to get compromised, *without requiring the system to be in the steady state*. The latter represents the probability that a node/computer is compromised in the steady state. Although it is hard to obtain closed-form expressions for these metrics, we manage to attain some analytic results centered on them. For example, we show that when the defense is highly effective, we can obtain some asymptotic result on computing the distribution of time-to-compromise with much less information (i.e., the mean of a certain distribution rather than the distribution); when the defense is poor, certain easier-to-obtain bounds are tight and can be used for decision-making purpose. This hints that security of both highly effectively and poorly defended networked systems would be easier to analyze than security of networked systems whose defense resides in the middle of the spectrum. To the best of our knowledge, this insight was not known until now.

### 1.2 The Science

This paper falls into the category of "mathematically modeling and analyzing security of networked systems," which is a core problem in the emerging Science of Security, because it aims to understand security from a holistic (or whole-system) perspective, rather than from the point of view of individual computers or building-blocks (e.g., protocols, mechanisms). This line of research is much needed, but little investigated perhaps because of the difficulties it imposes. As shown in the paper, such research can lead to *security metrics*, which can be used to *quantify* (and therefore *compare*) security of networked systems in a principled fashion. For example, in risk management, one would need to know the probability each computer is compromised at some point in time. Such probabilities may be based on the steady-state compromise probabilities of the nodes (i.e., "average case"-based decision-making) or based on the upper-bounds of the steady-state (or transient-state) compromise probabilities of the nodes (i.e., "worst-case"-based decision-making). Other applications of such probabilities include: The defender can deploy an appropriate threshold/proactive cryptosystem to tolerate compromised nodes. In Byzantine Agreement scheme, it is often assumed that no-more-than-one-third of the nodes are faulty/compromised; this assumption can be (in)validated by using the (upper bounds of) steady-state or transient-state compromise probabilities of the nodes.

The paper is organized as follows. Section 2 describes the new approach and the resulting model. Section 3 analyzes the model. Section 4 reviews related prior work. Section 5 discusses the limitations of the model. Section 6 concludes the paper. Proofs are deferred to the Appendix. The following table summarizes the main notations used in the paper:

| | |
|---|---|
| $P(\cdot), E[\cdot]$ | the probability and expectation functions |
| $G$ | $G = (V, E)$ is attack-defense structure graph, where $V$ represents computers and $E$ represents the "direct attack" relation |
| $\deg(v)$ | (in-)degree of node $v \in V$ in $G = (V, E)$ |
| $c$ | $c = (c_1, c_2)$, where $c_1, c_2 > 0$ are defense capabilities against push- and pull-based attacks |
| $J_v$ | random variable (local environment) abstracting $v$'s compromised neighbors |
| $\pi_{v,r}$ | probability mass function of $J_v$ |
| $J_{v,i}$ | number of compromised neighbors of $v \in V$ after the $(i-1)$th recovery, $i = 1, 2, \ldots$ |
| $\Theta_v$ | random variable (global environment) abstracting pull-based attacks against node $v$ |
| $H_v$ | distribution function of $\Theta_v$ |
| $T_{v,c}$ | *time-to-compromise* metric |
| $q_{v,c}(t)$ | distribution of $T_{v,c}$, $q_{v,c}(t) = P(T_{v,c} \leq t)$ |
| $\tilde{q}_{v,c}(t)$ | distribution of $T_{v,c}$ as $c_1, c_2 \to \infty$ |
| $T_{v,c,i}$ | the random time it takes for $v$ to change from secure to compromised state for the $i$th time |
| $R_v$ | the time it takes for $v \in V$ to change from the compromised state to the secure state |
| $R_{v,i}$ | the time it takes for $v$ to change from compromised to secure for the $i$th time |
| $p_{v,c}$ | *steady-state compromise probability* metric |
| $F_{i,r}^{(1)}, G_{i,r}^{(1)}$ | distributions of $X_i^{(1)}(r)$ and $Y_i^{(1)}(r)$ w.r.t. $r$ |
| $F_{i,\theta}^{(2)}, G_{i,\theta}^{(2)}$ | distributions of $X_i^{(2)}(\theta)$ and $Y_i^{(2)}(\theta)$ w.r.t. $\theta$ |

## 2. THE NEW APPROACH AND MODEL

### 2.1 The Attack-Defense Structure Abstraction

This abstraction has been used, implicitly or explicitly, to describe attacks and defenses in the past years (see, for example, [6, 5, 15, 25, 30, 27, 28]). Specifically, computers in a cyber system (e.g., an enterprise network and its associated information systems) can be abstracted as nodes (i.e., vertices) in terms of graph-theoretic models. We consider two major classes of cyber attacks:

- Push-based attacks: They are launched by malwares from the compromised computers against the vulnerable ones, by attempting to actively infect them.

- Pull-based attacks: These attacks are launched through mechanisms such as the drive-by-download attack (i.e., a vulnerable computer gets compromised because the user accessed a malicious website) and the insider attack that an authorized user intentionally executes a malicious software program on a computer. A computer compromised by a pull-based attack can further launch push-based attacks.

These two classes of attacks have been investigated by [14, 30, 15, 27], but using different approaches (see Section 4). Push-based attacks formulate an *attack-structure*, which is a graph $G = (V, E)$, where $V$ is the set of nodes (i.e., computers) and $E$ represents the *direct attack* relation that $(u, v) \in E$ means $u \in V$ can attack $v \in V$ directly (i.e., without using "stepping stones"). In other words, the *direct attack* relation imposes a graph structure $G$, which can have an arbitrary topology. For example, if a web server cannot launch push-based attacks but can be abused to host/launch pull-based attacks, the web server computer is not a node in $V$, but is part of the abstract "global environment" — the source of pull-based attacks. However, if the web server can also launch push-based attacks, then it is a node in $V$ and at the same time a part of the abstract "global environment." At any point in time, a node $v \in V$ is either compromised, or secure (but vulnerable).

In parallel to the *attack-structure*, there exists a *defense-structure* that represents how the defense takes place. In this paper we consider the following popular classes of defense mechanisms (which enforce security policies):

- Preventing known attacks: The defender uses (e.g.) network-based and/or host-based firewall and Intrusion Detection System (IDS) to filter known attacks. We use a pair of parameters, denoted by $(c_1, c_2)$, to denote the capability for filtering push-based and pull-based attacks that target at $v \in V$, respectively.

- Detecting and curing successful attacks: The defender uses (e.g.) anti-malware like mechanisms to detect malware infections and clean up the compromised computers. We use random variable $R_{v,i}$ to represent the time to detect and cure a compromised node $v \in V$ for the $i$th time.

As a result, the aforementioned attack structure $G = (V, E)$ naturally becomes the *attack-defense structure*, which is extended to accommodate the defense capabilities via parameters associated to the nodes and/or edges.

**Remark**. The above abstraction is sufficient for the purpose of the present paper (i.e., characterization study). It is an orthogonal research problem to obtain actual attack-defense structures and parameters for networked systems of interest.

### 2.2 The New Approach

We start with an illustration of attacks via Figure 1. Node $v \in V$ can be attacked by push-based attacks that are launched by $v$'s five compromised (incoming) neighbors in the attack-defense structure $G = (V, E)$. Note that our model and results are equally applicable to both directed and undirected attack-defense structures. Node $v \in V$ also can be attacked by pull-based attacks. The model has two aspects: specifying when $v$ is attacked (i.e., arrival of attacks), and specifying when $v$ is successfully attacked. This separation between *arrival of attack* and *success of attack* makes the resulting model more "native" than previous models for the same purpose [14, 30, 15, 27], meaning that the model is closer to real-life data that contains both unsuccessful and successful attacks.

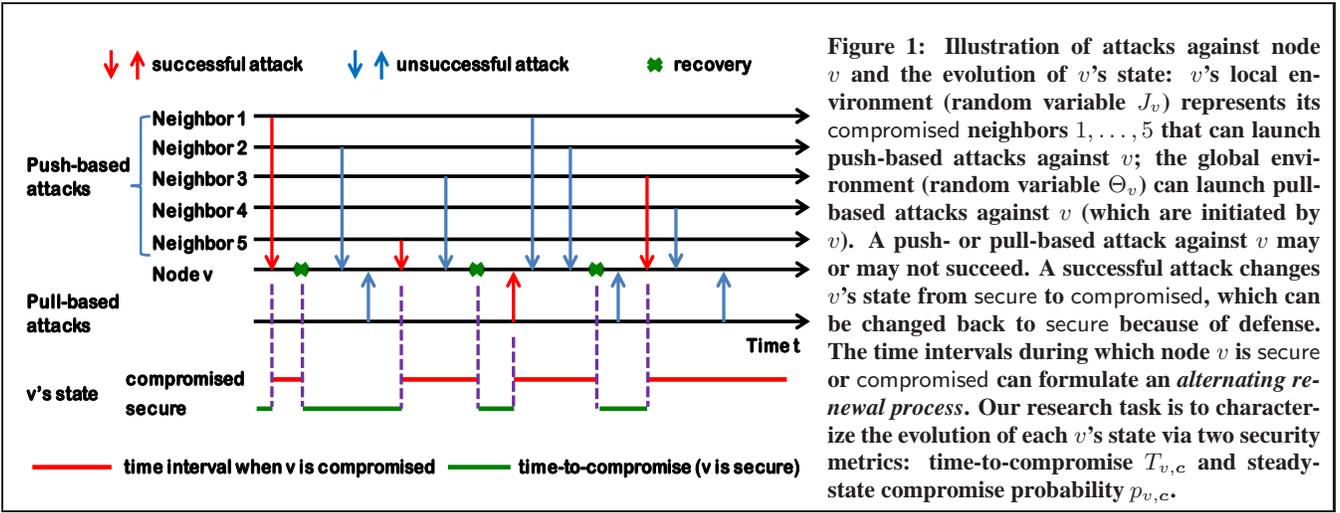

Figure 1: Illustration of attacks against node $v$ and the evolution of $v$'s state: $v$'s local environment (random variable $J_v$) represents its compromised **neighbors** $1, \ldots, 5$ that can launch push-based attacks against $v$; the global environment (random variable $\Theta_v$) can launch pull-based attacks against $v$ (which are initiated by $v$). A push- or pull-based attack against $v$ may or may not succeed. A successful attack changes $v$'s state from secure to compromised, which can be changed back to secure because of defense. The time intervals during which node $v$ is secure or compromised can formulate an *alternating renewal process*. Our research task is to characterize the evolution of each $v$'s state via two security metrics: time-to-compromise $T_{v,c}$ and steady-state compromise probability $p_{v,c}$.

**Specifying when a node is attacked (i.e., arrival of attack)**. We model push-based attacks against node $v$ via point process:

$$\left\{(X_i^{(1)}(J_v), Y_i^{(1)}(J_v))\right\}, \quad i = 0, 1, 2 \ldots,$$

where random variable $J_v$ represents the push-based local attack environment — the random number of $v$'s compromised (incoming) neighbors with support $\{0, \ldots, \deg(v)\}$ and $\deg(v)$ being the (in-)degree of $v$, random variable $X_i^{(1)}(J_v)$ represents the magnitude (i.e., sophistication or power) of the $i$th push-based attack against $v$ with $X_0^{(1)}(J_v) = 0$, and random variable $Y_i^{(1)}(J_v)$ represents the attack inter-arrival time between the $(i-1)$th and the $i$th push-based attacks against $v$ with $Y_0^{(1)}(J_v) = 0$.

Similarly, we model pull-based attacks via point process:

$$\left\{(X_i^{(2)}(\Theta_v), Y_i^{(2)}(\Theta_v))\right\}, i = 1, 2 \ldots,$$

where random variable $\Theta_v$ represents node $v$'s global pull-based attack environment (e.g., the extent of malicious websites presence in cyberspace), random variable $X_i^{(2)}(\Theta_v)$ represents the magnitude (i.e., sophistication or power) of the $i$th pull-based attack against $v$ with $X_0^{(2)}(\Theta_v) = 0$, and random variable $Y_i^{(2)}(\Theta_v)$ represents the attack inter-arrival time between the $(i-1)$th and the $i$th pull-based attacks against $v$ with $Y_0^{(2)}(\Theta_v) = 0$.

**Specifying when a node is compromised**. Having specified when an attack is launched against a secure node $v$, we now specify when $v$ becomes compromised. Let $c_1 > 0$ and $c_2 > 0$ abstract the defense capabilities in filtering push-based and pull-based attacks, respectively. A secure node $v$ becomes compromised when the magnitude of a push-based attack exceeds $c_1$, or when the magnitude of a pull-based attack exceeds $c_2$. The two defense capabilities accommodate defenses, including network-based and host-based.

**Specifying when a node recovers**. Because of defense (e.g., malware or intrusion detection), a compromised node $v$ becomes secure after a random time $R_{v,i}$ for the $i$th time.

**Remark**. Random variables $X_i^{(1)}(J_v)$ for $i = 1, 2, \ldots$ accommodate the adaptiveness of push-based attack magnitudes against node $v$ because they depend on the same random variable (i.e., local environment) $J_v$. Random variables $Y_i^{(1)}(J_v)$ for $i = 1, 2, \ldots$ accommodate the adaptiveness of push-based attack inter-arrival times. For example, a greater $J_v$ (in a stochastic sense) would imply a severer local environment, namely more intense attacks (i.e., smaller attack inter-arrival times) and/or greater attack magnitudes (i.e., more powerful attacks). Similarly, random variables $X_i^{(2)}(\Theta_v)$ for $i = 1, 2, \ldots$ accommodate the adaptiveness of pull-based attacks (e.g., different exploits targeting at different vulnerabilities in the browser/computer) against $v$, because they depend on the same random variable (i.e., global environment) $\Theta_v$. Random variables $Y_i^{(2)}(\Theta_v)$ for $i = 0, 1, 2, \ldots$ accommodate the adaptiveness of inter-arrival times between pull-based attacks against $v$. We will describe how the assumptions that we will make (for the sake of analytic tractability) and the results accommodate adaptiveness.

### 2.3 Security Metrics

**Metric 1: Time-to-compromise $T_{v,c}$ where $c = (c_1, c_2)$**. As illustrated in Figure 1, this metric, denoted by $T_{v,c}$, captures the time it takes for node $v$ to change from the secure state to the compromised state. Since $T_{v,c}$ is a random variable in general, we consider its distribution $q_{v,c}(t) = P(T_{v,c} \leq t)$. Specifically, for given local environment $J_v = r$ and global environment $\Theta_v = \theta$, where $0 \leq r \leq \deg(v)$ and $\theta > 0$ (e.g., the portion of compromised websites), let $N_r^{(1)}(t)$ and $N_\theta^{(2)}(t)$ be the counting processes associated to sequences $\{Y_i^{(1)}(r), i \geq 0\}$ and $\{Y_i^{(2)}(\theta), i \geq 0\}$, respectively. We respectively define the historical maximum (up to time $t$) of the push- and pull-based attack magnitudes as

$$\begin{aligned} M_r^{(1)}(t) &= \vee_{i=0}^{N_r^{(1)}(t)} X_i^{(1)}(r), \\ M_\theta^{(2)}(t) &= \vee_{i=0}^{N_\theta^{(2)}(t)} X_i^{(2)}(\theta), \end{aligned} \quad (1)$$

where "$\vee$" means taking the maximum. The time-to-compromise for node $v$, due to push-based attacks launched from the given $r$ compromised neighbors and due to pull-based attacks from the given global environment $\theta$, can be respectively defined as

$$\begin{aligned} T_{c_1}^{(1)}(r) &= \inf\{t : M_r^{(1)}(t) > c_1\}, \\ T_{c_2}^{(2)}(\theta) &= \inf\{t : M_\theta^{(2)}(t) > c_2\}. \end{aligned} \quad (2)$$

Therefore, the time-to-compromise for node $v$ is

$$T_c(r, \theta) = T_{c_1}^{(1)}(r) \wedge T_{c_2}^{(2)}(\theta), \quad (3)$$

where "∧" means taking the minimum.

The above reasoning is for given $\boldsymbol{c} = (c_1, c_2)$ and $(J_v = r, \Theta_v = \theta)$. For random environments $(J_v, \Theta_v)$, we obtain the time-to-compromise for node $v$ as

$$T_{v,\boldsymbol{c}} \equiv T_{\boldsymbol{c}}(J_v, \Theta_v) = T_{c_1}^{(1)}(J_v) \wedge T_{c_2}^{(2)}(\Theta_v), \quad (4)$$

which is the mixture of $T_{\boldsymbol{c}}(r, \theta)$, with respect to $J_v$ and $\Theta_v$.

Note that the above discussion does not rely on any assumption, and applies even if the system is not in the steady state. The discussion holds under the assumptions discussed below.

**Metric 2: Steady-state compromise probability** $p_{v,\boldsymbol{c}}$. As illustrated in Figure 1, the state of $v \in V$ alternates with time. For given defense capabilities $\boldsymbol{c} = (c_1, c_2)$, the state of $v \in V$ at time $t$ can be seen as Bernoulli random variable $X_{v,\boldsymbol{c}}(t)$, where $X_{v,\boldsymbol{c}}(t) = 1$ means $v$ is compromised and $X_{v,\boldsymbol{c}}(t) = 0$ means $v$ is secure. Let $p_{v,\boldsymbol{c}}(t) = \mathrm{P}(X_{v,\boldsymbol{c}}(t) = 1)$ be the probability that $v$ is compromised at time $t$. We aim to obtain the probability that $v$ is compromised in the steady state: $p_{v,\boldsymbol{c}} = \lim_{t \to \infty} p_{v,\boldsymbol{c}}(t)$.

## 3. ANALYZING THE MODEL

**Preliminaries**. Our analysis is centered on the two security metrics, while using the following definition [3, 16].

DEFINITION 1. *Let $\bar{F} = 1 - F$ be the survival function of distribution $F$. $F$, or the corresponding random variable $Z$, is said to be:*

(i) *NBU (New Better Than Used): This property says $\bar{F}(z_1 + z_2) \leq \bar{F}(z_1)\bar{F}(z_2)$ for $z_1, z_2 \geq 0$, namely that the conditional survival probability satisfies $\mathrm{P}(Z > z_1 + z_2 | Z > z_1) \leq \mathrm{P}(Z > z_2)$.*

(ii) *NBUE (New Better Than Used in Expectation): This property says $\int_z^\infty \bar{F}(x)\,dx \leq \mathrm{E}[Z]\bar{F}(z)$ for $z \geq 0$, namely that the conditional expectation satisfies $\mathrm{E}[Z - z | Z > z] \leq \mathrm{E}[Z]$.*

We will consider attack inter-arrival time that exhibits such memory properties. This goes beyond the common practice of assuming that the attack inter-arrival time follows the exponential distribution, whose memoryless property warrants analytic tractability. We consider NBU or NBUE attack inter-arrival time, which contains the exponential, Weibull, Pareto and Gamma distributions as special cases. Note that NBU implies NBUE, but not vice versa. As we will elaborate later, the NBU/NBUE memory property accommodates a kind of adaptiveness or dependence that is different from what is accommodated by the random environment technique.

### 3.1 Analyzing Time-to-Compromise Metric $T_{v,\boldsymbol{c}}$

#### 3.1.1 Distribution function $q_{v,\boldsymbol{c}}(t)$

For the sake of tractability, we propose to retain the adaptiveness that can be accommodated by local environment $J_v$ and global environment $\Theta_v$, but assume independence between the relevant random variables *conditioned on given specific environments* $J_v = r$ and $\Theta_v = \theta$. This is a popular method for coping with dependence in probability and statistics.

ASSUMPTION 1. (a) *For any $v \in V$ and for given local environment $J_v = r$, where $0 \leq r \leq \deg(v)$:*

(i) $\{X_i^{(1)}(r), i \geq 1\}$ *is an independent sequence;*

(ii) $\{Y_i^{(1)}(r), i \geq 1\}$ *is an independent sequence;*

(iii) $\{X_i^{(1)}(r), i \geq 1\}$ *and $\{Y_i^{(1)}(r), i \geq 1\}$ are independent of each other.*

(b) *For any $v \in V$ and for given global environment $\Theta_v = \theta$:*

(i) $\{X_i^{(2)}(\theta), i \geq 1\}$ *is an independent sequence;*

(ii) $\{Y_i^{(2)}(\theta), i \geq 1\}$ *is an independent sequence;*

(iii) $\{X_i^{(2)}(\theta), i \geq 1\}$ *and $\{Y_i^{(2)}(\theta), i \geq 1\}$ are independent of each other.*

(c) *For any $v \in V$, for given local environment $J_v = r$, and for given global environment $\Theta_v = \theta$: $\{(X_i^{(1)}(r), Y_i^{(1)}(r), i \geq 1\}$ and $\{(X_i^{(2)}(\theta), Y_i^{(2)}(\theta)), i \geq 1\}$ (i.e., push-based attacks and pull-based attacks) are independent of each other. For technical simplicity, we further assume that $J_v$ and $\Theta_v$ are independent of each other. These collectively imply the following property: $\{(X_i^{(1)}(J_v), Y_i^{(1)}(J_v)), i \geq 1\}$ and $\{(X_i^{(2)}(\Theta_v), Y_i^{(2)}(\Theta_v)), i \geq 1\}$ are independent of each other.*

**How does Assumption 1 accommodate adaptiveness of attacks?** First, the above (a).(i) does not imply that the magnitudes of push-based attacks $\{X_i^{(1)}(J_v), i \geq 1\}$, under random environment $J_v$ (rather than a given $J_v = r$ compromised neighbors), are independent. In contrast, $\{X_i^{(1)}(J_v), i \geq 1\}$ are dependent because of the following [21]: Suppose $X_i^{(1)}(r)$ is increasing or decreasing (i.e., it is the monotonicity that matters) in $r$ for any $i \geq 1$ in the *likelihood ratio order* [20], which is not ruled out by the above (a).(i). Then, $f_{r+1}(x)/f_r(x)$ is increasing or decreasing, where $f_{r+1}(x)$ and $f_r(x)$ are density functions of $X_i^{(1)}(r+1)$ and $X_i^{(1)}(r)$, respectively. For any $s \leq s'$, we have

$$\mathrm{P}\left(X_{i+1}^{(1)}(J_v) \geq x | X_i^{(1)}(J_v) = s\right)$$
$$\leq \mathrm{P}\left(X_{i+1}^{(1)}(J_v) \geq x | X_i^{(1)}(J_v) = s'\right).$$

This means that a larger magnitude $s'$ (i.e., a failed sophisticated attack) is more likely followed by another (more) sophisticated attack, namely that the attacker can adaptively increases its attack power (until an attack succeeds). The fundamental reason for that $\{X_i^{(1)}(r), i \geq 1\}$ for a given $r$ is independent but $\{X_i^{(1)}(J_v), i \geq 1\}$ is dependent is exactly that $X_i^{(1)}(r)$ is increasing or decreasing in $r$ for any $i \geq 1$.

Second, the independence between the attack inter-arrival times in (a).(ii) is for *given* local environment $J_v = r$; whereas attack inter-arrival times between push-based attacks $Y_i^{(1)}(J_v)$ for $i \geq 1$, under random environment $J_v$, can be dependent. Suppose $Y_i^{(1)}(r)$ is decreasing (or increasing) in $r$ for any $i \geq 1$ in the likelihood ratio order. It is also known [21] that for $s \leq s'$

$$\mathrm{P}\left(Y_{i+1}^{(1)}(J_v) \leq y | Y_i^{(1)}(J_v) = s\right)$$
$$\geq \mathrm{P}\left(Y_{i+1}^{(1)}(J_v) \leq y | Y_i^{(1)}(J_v) = s'\right),$$

This means that a smaller inter-arrival time $s$ (i.e., a failed intense attack) is more likely followed by another (more) intense attack, namely that the attacker can adaptively reduce its attack time (until an attack succeeds).

Third, the above (a).(iii) does not imply $\{X_i^{(1)}(J_v), i \geq 1\}$ and $\{Y_i^{(1)}(J_v), i \geq 1\}$ are independent. Rather, they are dependent because both magnitudes and attack inter-arrival times are dependent on the local environment $J_v$.

The preceding discussion focuses on Assumption 1.(a) and is equally applicable to Assumption 1.(b).

**When would (not) Assumption 1 hold?** Assumption 1 can be violated by full-fledged *adaptive* and *coordinated* attacks. This means that our model accommodates a certain degree of adaptiveness. Nevertheless, Assumption 1(c) would often hold because pull-based attacks are essentially governed by user behaviors (e.g., how often a user accesses malicious websites), while pull-based attacks are not. Moreover, the software vulnerabilities that are exploited by pull-based attacks are often different from the software vulnerabilities that are exploited by push-based attacks.

**Characterizing distribution function $q_{v,c}(t)$ under Assumption 1.** Denote by $F_{i,r}^{(1)}(\cdot)$ and $F_{i,\theta}^{(2)}(\cdot)$ respectively the distribution functions of $X_i^{(1)}(r)$ and $X_i^{(2)}(\theta)$, by $G_{i,r}^{(1)}(\cdot)$ and $G_{i,\theta}^{(2)}(\cdot)$ respectively the distribution functions of $Y_i^{(1)}(r)$ and $Y_i^{(2)}(\theta)$, where $0 \leq r \leq \deg(v)$ and $\theta > 0$. For fixed defense capabilities $c_1, c_2 > 0$, from Eqs. (1) and (2) as well as Assumption 1(a) we have:

$$\begin{aligned}
&\mathrm{P}\left(T_{c_1}^{(1)}(r) > t\right) \\
&= \mathrm{P}\left(\vee_{i=0}^{N_r^{(1)}(t)} X_i^{(1)}(r) < c_1\right) \\
&= \sum_{m=0}^{\infty}\left[\mathrm{P}\left(\vee_{i=0}^{N_r^{(1)}(t)} X_i^{(1)}(r) < c_1 | N_r^{(1)}(t) = m\right) \cdot \right.\\
&\qquad\left. \mathrm{P}\left(N_r^{(1)}(t) = m\right)\right] \\
&= \sum_{m=0}^{\infty} \prod_{i=1}^{m} F_{i,r}^{(1)}(c_1) \mathrm{P}\left(N_r^{(1)}(t) = m\right). \quad (5)
\end{aligned}$$

Similarly, from Assumption 1(b) we can obtain

$$\mathrm{P}\left(T_{c_2}^{(2)}(\theta) > t\right) = \sum_{m=0}^{\infty}\prod_{i=1}^{m} F_{i,\theta}^{(2)}(c_2) \mathrm{P}\left(N_\theta^{(2)}(t) = m\right), \quad (6)$$

where we define $\prod_{i=1}^{0} a_i = 1$ for any $a_i$. From Eq. (3) and Assumption 1(c), we have

$$\begin{aligned}
\mathrm{P}(T_c(r,\theta) > t) &= \mathrm{P}\left(T_{c_2}^{(1)}(r) > t, T_{c_2}^{(2)}(\theta) > t\right) \\
&= \left[\sum_{m=0}^{\infty}\prod_{i=1}^{m} F_{i,r}^{(1)}(c_2) \mathrm{P}\left(N_r^{(1)}(t) = m\right)\right] \cdot \\
&\quad \left[\sum_{m=0}^{\infty}\prod_{i=1}^{m} F_{i,\theta}^{(2)}(c_2) \mathrm{P}\left(N_\theta^{(2)}(t) = m\right)\right].
\end{aligned}$$

Let $\pi_{v,r}$ be the probability mass function of $J_v$ and $H_v(\cdot)$ be the distribution function of $\Theta_v$. Since $T_{v,c}$ is the mixture of $T_c(r,\theta)$ with respect to $J_v$ and $\Theta_v$, we have

$$\begin{aligned}
&q_{v,c}(t) \\
&= \mathrm{P}(T_{v,c} \leq t) \\
&= 1 - \mathrm{E}\left[\mathrm{E}\left[I(T_{v,c} > t) | J_v, \Theta_v\right]\right] \\
&= 1 - \mathrm{E}\left[\mathrm{E}\left[I\left(T_{c_1}^{(1)}(J_v) > t, T_{c_2}^{(2)}(\Theta_v) > t\right) | J_v, \Theta_v\right]\right] \\
&= 1 - \mathrm{E}\left[\sum_{m=0}^{\infty}\prod_{i=1}^{m} F_{i,J_v}^{(1)}(c_1) \mathrm{P}\left(N_{J_v}^{(1)}(t) = m\right)\right] \cdot \\
&\quad \mathrm{E}\left[\sum_{m=0}^{\infty}\prod_{i=1}^{m} F_{i,\Theta_v}^{(2)}(c_2) \mathrm{P}\left(N_{\Theta_v}^{(2)}(t) = m\right)\right] \\
&= 1 - \sum_{r=0}^{d(v)} \pi_{v,r} \sum_{m=0}^{\infty}\prod_{i=1}^{m} F_{i,r}^{(1)}(c_1) \mathrm{P}\left(N_r^{(1)}(t) = m\right) \cdot \\
&\quad \int_0^{\infty} \sum_{m=0}^{\infty}\prod_{i=1}^{m} F_{i,\theta}^{(2)}(c_2) \mathrm{P}\left(N_\theta^{(2)}(t) = m\right) dH_v(\theta). \quad (7)
\end{aligned}$$

### 3.1.2 An upper bound for $q_{v,c}(t)$: $q_{v,c}(t)^+$

Since we cannot attain simple closed-form expression of $q_{v,c}(t)$ as shown in Eq. (7), we aim to bound it by making the following Assumption 2.

ASSUMPTION 2. *(a) The same as Assumption 1(a).*

*(b) The same as Assumption 1(b).*

*(c) For any $v \in V$ and given environments $J_v = r$ and $\Theta_v = \theta$, $\left\{\left(X_i^{(1)}(r), Y_i^{(1)}(r)\right), i \geq 1\right\}$ are independently and identically distributed samples of $\left\{(X^{(1)}(r), Y^{(1)}(r))\right\}$, and $\left\{\left(X_i^{(2)}(\theta), Y_i^{(2)}(\theta)\right), i \geq 1\right\}$ are independently and identically distributed samples of $\left\{(X^{(2)}(\theta), Y^{(2)}(\theta))\right\}$.*

Now we present an upper bound for $q_{v,c}(t)$, denoted by $q_{v,c}^+(t)$. Denote by $F_r^{(1)}(t)$, $G_r^{(1)}(t)$, $F_\theta^{(2)}(t)$ and $G_\theta^{(2)}(t)$ the distributions of $X^{(1)}(r)$, $Y^{(1)}(r)$, $X^{(2)}(\theta)$ and $Y^{(2)}(\theta)$, respectively. Under Assumption 2, from Eq. (7), we have

$$q_{v,c}(t) = 1 - \mathrm{E}\left[(F_{J_v}^{(1)}(c_1))^{N_{J_v}^{(1)}(t)}\right] \mathrm{E}\left[(F_{\Theta_v}^{(2)}(c_2))^{N_{\Theta_v}^{(2)}(t)}\right]. \quad (8)$$

PROPOSITION 1. *(upper bound of $q_{v,c}(t)$) Suppose Assumption 2 holds, and $Y^{(1)}(r)$ and $Y^{(2)}(\theta)$ have the NBU property for any given local environment $J_v = r$ and global environment $\Theta_v = \theta$. We have*

$$\begin{aligned}
&q_{v,c}^+(t) \\
&= 1 - \sum_{r=0}^{\deg(v)} \pi_{v,r} [\bar{G}_r^{(1)}(t)]^{\bar{F}_r^{(1)}(c_1)} \int_0^{\infty} [\bar{G}_\theta^{(2)}(t)]^{\bar{F}_\theta^{(2)}(c_2)} \, dH_v(\theta).
\end{aligned}$$

**How does Proposition 1 (i.e., Assumption 2 and NBU attack inter-arrival times) accommodate adaptiveness of attacks?** On one hand, Assumption 2 is slightly stronger than Assumption 1, meaning that the adaptiveness accommodated by Assumption 2 may be slightly weaker than the adaptiveness accommodated by Assumption 1. On the other hand, under Assumption 2, $\{Y_i^{(1)}(r), i \geq 1\}$ are independently and identically distributed according to distribution $Y^{(1)}(r)$, whose NBU property brings a certain other kind of adaptiveness that is not accommodated by Assumption 1. Specifically, the NBU property says

$$\mathrm{P}\left(Y^{(1)}(r) > z_1 + z_2 | Y^{(1)}(r) > z_1\right) \leq \mathrm{P}\left(Y^{(1)}(r) > z_2\right).$$

By treating the attack events as a stationary point process with attack inter-arrival times that exhibit the NBU property, we see that for any $z_1, z_2 \geq 0$,

$$\mathrm{P}\left(Y_{i+1}^{(1)}(r) > z_1 + z_2 | Y_{i+1}^{(1)}(r) > z_1\right) \leq \mathrm{P}\left(Y_i^{(1)}(r) > z_2\right).$$

That is, the extra-waiting time for the $(i+1)$th attack under the condition that the attack has not arrived after some time is smaller than the waiting time for the $i$th attack in the stochastic sense. It is worthwhile to highlight that the adaptiveness accommodated by Assumption 1 is caused by that $Y_i^{(1)}(r)$ is decreasing or increasing in $r$ in the likelihood ratio order; whereas, the adaptiveness accommodated by NBU is caused by the "memory" property of the distribution of $Y^{(1)}(r)$ for *fixed* $r$.

### 3.1.3 Expectation function $\mathrm{E}[T_{v,c}]$ and its lower bound

We present two lower bounds based on different assumptions.

**One lower bound $\mathrm{E}[T_{v,\mathbf{c}}]^-$ under Assumption 2 and NBUE attack inter-arrival times**. From (7), it can be shown that

$$\begin{aligned}
\mathrm{E}[T_{v,\mathbf{c}}] &= \int_0^\infty (1 - q_{v,\mathbf{c}}(t)) dt \\
&= \mathrm{E}\left[\sum_{\ell=0}^\infty \sum_{m=0}^\infty \prod_{i=1}^m F_{i,J_v}^{(1)}(c_1) \prod_{j=1}^m F_{j,\Theta_v}^{(2)}(c_2) \right. \\
&\quad \left. \int_0^\infty p\left(N_{J_v}^{(1)}(t) = \ell\right) \mathrm{P}\left(N_{\Theta_v}^{(2)}(t) = m\right) dt\right].
\end{aligned}$$

Since we cannot derive simple closed-form expression for $\mathrm{E}[T_{v,\mathbf{c}}]$, we derive a lower bound of $\mathrm{E}[T_{v,\mathbf{c}}]$, denoted by $\mathrm{E}[T_{v,\mathbf{c}}]^-$, by considering the two classes of attacks separately.

From Eqs. (5) and (6) we have

$$\begin{aligned}
\mathrm{E}[T_{c_1}^{(1)}(r)] &= \sum_{m=0}^\infty \prod_{i=1}^m F_{i,r}^{(1)}(c_1) \int_0^\infty \mathrm{P}(N_r^{(1)}(t) = m) \, dt \\
&= \sum_{m=0}^\infty \prod_{i=1}^m F_{i,r}^{(1)}(c_1) \mathrm{E}[Y_{m+1}^{(1)}(r)], \quad (9)
\end{aligned}$$

and

$$\mathrm{E}[T_{c_2}^{(2)}(\theta)] = \sum_{m=0}^\infty \prod_{i=1}^m F_{i,\theta}^{(2)}(c_2) \mathrm{E}[Y_{m+1}^{(2)}(\theta)]. \quad (10)$$

Under Assumption 2, the two expectations in Eqs. (9) and (10) can be expressed as

$$\mathrm{E}[T_{c_1}^{(1)}(r)] = \frac{\mathrm{E}[Y^{(1)}(r)]}{\bar{F}_r^{(1)}(c_1)}, \quad \mathrm{E}[T_{c_2}^{(2)}(\theta)] = \frac{\mathrm{E}[Y^{(2)}(\theta)]}{\bar{F}_\theta^{(2)}(c_2)}. \quad (11)$$

PROPOSITION 2. (lower bound $\mathrm{E}[T_{v,\mathbf{c}}]^-$) *Suppose Assumption 2 holds, and $Y^{(1)}(r)$ and $Y^{(2)}(\theta)$ have the NBUE property for any given local environment $J_v = r$ and global environment $\Theta_v = \theta$. We have*

$$\mathrm{E}[T_{v,\mathbf{c}}]^- = \sum_{r=0}^{\deg(v)} \int_0^\infty \pi_{v,r} \left(\frac{\bar{F}_r^{(1)}(c_1)}{\mathrm{E}[Y^{(1)}(r)]} + \frac{\bar{F}_\theta^{(2)}(c_2)}{\mathrm{E}[Y^{(2)}(\theta)]}\right)^{-1} dH_v(\theta).$$

**How does Proposition 2 (i.e., Assumption 2 and NBUE attack inter-arrival times) accommodate adaptiveness of attacks?** As mentioned above, the adaptiveness accommodated by Assumption 2 may be slightly weaker than the adaptiveness accommodated by Assumption 1. On the other hand, under Assumption 2, the NBUE property brings a certain other degree of adaptiveness. Specifically, the NBUE property says

$$\mathrm{E}\left[Y^{(1)}(r) - z | Y^{(1)}(r) > z\right] \leq \mathrm{E}\left[Y^{(1)}(r)\right].$$

Similarly, by treating the attack events as a stationary point process with all attack inter-arrival times having a common marginal distribution, the NBUE property implies that for any $z \geq 0$,

$$\mathrm{E}[Y_{i+1}^{(1)}(r) - z | Y_{i+1}^{(1)}(r) > z] \leq \mathrm{E}[Y_i^{(1)}(r)] \text{ for } i \geq 1.$$

That is, the expected waiting time for the $i$th attack is greater than the expected extra-waiting time under the condition that the $(i+1)$th attack has not arrived within time $z$. It is also worth mentioning that the adaptiveness accommodated by Assumption 1 is caused by that $Y_i^{(1)}(r)$ is decreasing or increasing in $r$ in the likelihood ratio order; whereas, the adaptiveness accommodated by NBUE is caused by the "memory" property of the distribution of $Y^{(1)}(r)$.

**Another lower bound $\mathrm{E}[T_{v,\mathbf{c}}]^-$ under Assumption 3 and NBUE attack inter-arrival times**. The lower bound given in Proposition 2 is useful because it only requires information about the expectations of the attack inter-arrival time (rather than the entire distribution). In what follows, we establish another lower bound under a different assumption.

ASSUMPTION 3. *This assumption is the same as Assumption 1, except that Assumption 1(a).(i) and 1(b).(i) are respectively replaced by the following:*

*(a).(i) For any given local environment $J_v = r$, $\{X_i^{(1)}(r), i \geq 1\}$ is an independent and increasing sequence (in $i$) in the usual stochastic order sense.*

*(b).(i) For any global environment $\Theta_v = \theta$, and $\{X_i^{(2)}(\theta), i \geq 1\}$ is an independent and increasing sequence (in $i$) in the usual stochastic order sense.*

PROPOSITION 3. *Suppose Assumption 3 holds. Suppose $Y_i^{(1)}(r)$ and $Y_i^{(2)}(\theta)$ have the NBUE property for any given local environment $J_v = r$ and global environment $\Theta_v = \theta$. Suppose $\mathrm{E}[Y_i^{(1)}(r)]$ and $\mathrm{E}[Y_i^{(2)}(\theta)]$ are decreasing in $i$ for $i \geq 1$. We have*

$$\begin{aligned}
&\mathrm{E}[T_{v,\mathbf{c}}]^- \\
&= \sum_{r=0}^{\deg(v)} \pi_{v,r} \int_0^\infty (\mathrm{E}^{-1}[T_{c_1}^{(1)}(r)] + \mathrm{E}^{-1}[T_{c_2}^{(2)}(\theta)])^{-1} \, dH_v(\theta),
\end{aligned}$$

*where $\mathrm{E}[T_{c_1}^{(1)}(r)]$ and $\mathrm{E}[T_{c_2}^{(2)}(\theta)]$ are respectively given in Eqs. (9) and (10).*

**How does Proposition 3 accommodate adaptiveness of attacks?** First, Assumption 3 is slightly stronger than Assumption 1, meaning that the adaptiveness accommodated by Assumption 3 may be slightly weaker than the adaptiveness that is accommodated by Assumption 1. Second, the enhanced Assumption 3(a).(i) and 3(b).(i) accommodate a certain other kind of adaptiveness, namely

$$\begin{aligned}
\mathrm{P}\left(X_{i+1}^{(1)}(r) > s\right) &\geq \mathrm{P}\left(X_i^{(1)}(r) > s\right), \text{ and} \\
\mathrm{P}\left(X_{i+1}^{(1)}(\theta) > s\right) &\geq \mathrm{P}\left(X_i^{(1)}(\theta) > s\right),
\end{aligned}$$

which accommodate increasing magnitudes (or capabilities) of attacks. Third, the assumption that $\mathrm{E}[Y_i^{(1)}(r)]$ and $\mathrm{E}[Y_i^{(2)}(\theta)]$ are decreasing in $i$ for $i \geq 1$ further accommodates the following adaptiveness in attacks inter-arrive times:

$$\mathrm{E}\left[Y_{i+1}^{(1)}(r)\right] \leq \mathrm{E}\left[Y_i^{(1)}(r)\right] \text{ and } \mathrm{E}\left[Y_{i+1}^{(1)}(\theta)\right] \leq \mathrm{E}\left[Y_i^{(1)}(\theta)\right],$$

namely that attacks can get more intense.

### 3.1.4 Example: Numerical solution to $q_{v,\mathbf{c}}(t)$ and tightness of upper bound $q_{v,\mathbf{c}}^+(t)$ and lower bound $\mathrm{E}[T_{v,\mathbf{c}}]^-$

**On the tightness of upper bound $q_{v,\mathbf{c}}^+(t)$.** Consider a random node $v \in V$. Suppose $X^{(1)}(r)$, the random magnitude of any push-based attack launched from the local environment $J_v = r$ compromised neighbors, is Weibull random variable with shape parameter $\alpha$, scale parameter $1/r$, and distribution function

$$F_r^{(1)}(t) = 1 - e^{-(r^{-1}t)^\alpha}, \quad \alpha > 0, \, t > 0.$$

Suppose $Y^{(1)}(r)$, the inter-arrival time between any two consecutive push-based attacks, is Gamma random variable with shape parameter $\beta \geq 1$, scale parameter $r$, and distribution function

$$G_r^{(1)}(t) = \int_0^t \frac{r^\beta x^{\beta-1} e^{-rx}}{\Gamma(\beta)} \, dx, \quad \beta \geq 1, \, t > 0.$$

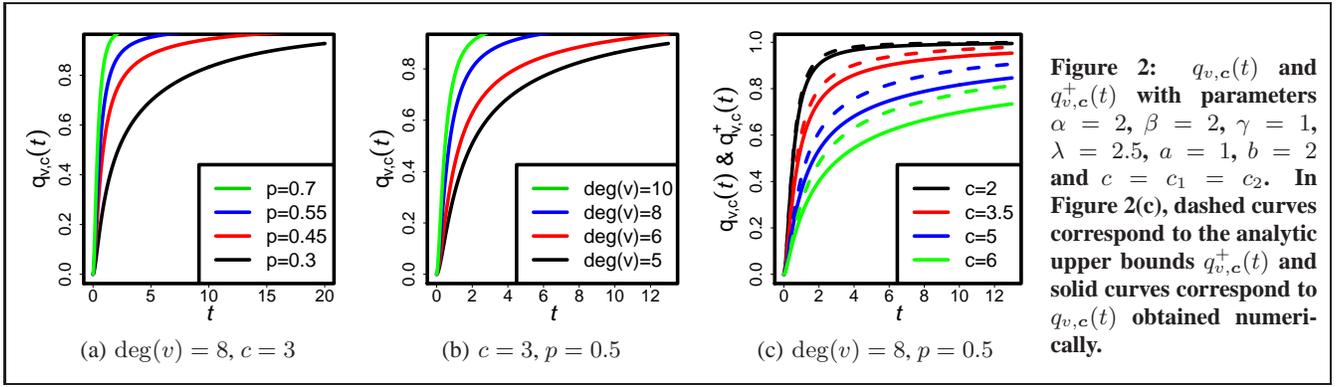

Figure 2: $q_{v,c}(t)$ and $q_{v,c}^+(t)$ with parameters $\alpha = 2$, $\beta = 2$, $\gamma = 1$, $\lambda = 2.5$, $a = 1$, $b = 2$ and $c = c_1 = c_2$. In Figure 2(c), dashed curves correspond to the analytic upper bounds $q_{v,c}^+(t)$ and solid curves correspond to $q_{v,c}(t)$ obtained numerically.

Note that if $\beta = 1$, the attack inter-arrival time becomes the *exponential distribution*, which is often assumed in many previous security models and justifies why our model is general. Suppose $J_v$ follows the Binomial distribution with parameter $p$. The probability that $v$ has $r$ compromised neighbors is:

$$\pi_{v,r} = \binom{\deg(v)}{r} p^r (1-p)^{(\deg(v)-r)}, \ r = 0, \ldots, \deg(v).$$

Given $\Theta_v = \theta$, suppose the magnitude $X^{(2)}(\theta)$ of any pull-based attack is Weibull random variable with shape parameter $\gamma$, scale parameter $1/\theta$, and distribution function

$$F_\theta^{(2)}(t) = 1 - e^{-(\theta^{-1}t)^\gamma}, \ \gamma > 0, \ t > 0.$$

Suppose the inter-arrival time $Y^{(2)}(\theta)$ between any two consecutive pull-based attacks is Gamma random variable with shape parameter $\lambda \geq 1$, scale parameter $\theta$, and distribution function

$$G_\theta^{(2)}(t) = \int_0^t \frac{\theta^\lambda x^{\lambda-1} e^{-\theta x}}{\Gamma(\lambda)} \, dx, \ \lambda \geq 1, \ t > 0.$$

Suppose environment $\Theta_v$ is uniformly distributed in $[a, b]$ (i.e., pull-based attacks are from a uniform environment), namely

$$H_v(\theta) = \frac{\theta - a}{b - a}, \ a \leq \theta \leq b, \ 0 \leq a < b.$$

Note that for all $r$ and $\theta$, $Y^{(1)}(r)$ and $Y^{(2)}(\theta)$ have the NBU property because their shape parameters $\beta \geq 1$ and $\lambda \geq 1$.

Figure 2 demonstrates how the parameters affect the $q_{v,c}(t)$ as specified in Eq. (8), where $c = c_1 = c_2$. Figure 2(a) shows how the local environment affects $q_{v,c}(t)$, where a greater $p$ means a severer environment (i.e., more compromised neighbors) and leads to a greater $q_{v,c}(t)$ or a smaller time-to-compromise. Figure 2(b) shows that the greater the node degree, the greater $q_{v,c}(t)$ or the smaller the time-to-compromise. Figure 2(c) plots the upper bound $q_{v,c}^+(t)$ (dashed curves) derived from Proposition 1 and the probability $q_{v,c}(t)$ obtained numerically according to Eq. (8). We observe that $q_{v,c}^+(t)$ is tight for small $c = c_1 = c_2$, meaning that $q_{v,c}^+(t)$ *is particularly useful in the scenario of less effectively defended networked systems* (e.g., $c = c_1 = c_2 \leq 2$); precisely mapping $c_1$ and $c_2$ to actual attacks can be based on cyber attack-defense experiments, which are orthogonal to the focus of the present paper. We also observe that a greater $c$ leads to a smaller $q_{v,c}(t)$ and a greater time-to-compromise $T_{v,c}$ (because a better defended networked system is relatively harder to penetrate into).

**On the tightness of lower bound** $\mathrm{E}[T_{v,c}]^-$. Figure 3 demonstrates the tightness of lower bound of $\mathrm{E}[T_{v,c}]^-$ as derived from Proposition 2. Figure 3(a) shows that $\mathrm{E}[T_{v,c}]^-$ is tight for $p = .2$ or relatively secure environment. Figure 3(b) shows that $\mathrm{E}[T_{v,c}]^-$ is tight for $p = .8$ or relatively more malicious environment. As such, $\mathrm{E}[T_{v,c}]^-$ could be used in place of $\mathrm{E}[T_{v,c}]$ for decision-making, which is useful because $\mathrm{E}[T_{v,c}]^-$ is easier to obtain (i.e., requires little information). It is also observed that both $\mathrm{E}[T_{v,c}]$ and $\mathrm{E}[T_{v,c}]^-$ increase in $c = c_1 = c_2$, since better defense (i.e., larger $c$) leads to greater time-to-compromise time. This further suggests that although it is hard to precisely compute security metrics for the case of less effectively defended networked systems (i.e., small $c_1$ and $c_2$), *the defender could use the easier-to-obtain bounds for decision-making purpose*.

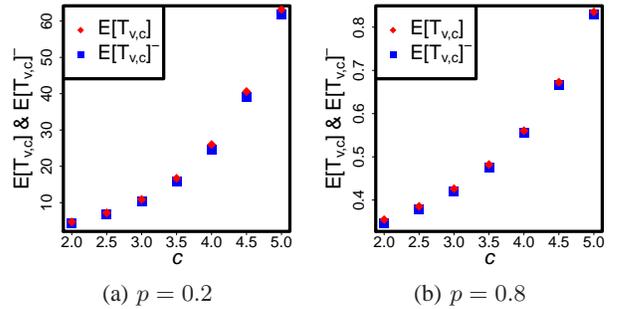

Figure 3: Tightness of $\mathrm{E}[T_{v,c}]^-$: $c = c_1 = c_2$, $\alpha = 2$, $\beta = 2$, $\gamma = 1$, $\lambda = 2.5$, $a = 1$, $b = 2$, $\deg(v) = 8$ and $p = .2, .8$. Since the easier-to-obtain lower bound is tight, the defender can use it in decision making.

As shown in Figures 2 and 3, both $q_{v,c}^+(t)$ and $\mathrm{E}[T_{v,c}]^-$ are tight when $c = c_1 = c_2$ is small (e.g., $c = 2$). This is no coincident, and can be explained by the following fact:

$$\mathrm{E}[T_{v,c}] = \int_0^\infty (1 - q_{v,c}(t)) \, dt \geq \int_0^\infty \left(1 - q_{v,c}^+(t)\right) dt$$
$$\geq \mathrm{E}[T_{v,c}]^-,$$

where the last inequality can be obtained from the proof of Proposition 2 (in the Appendix).

### 3.2 Asymptotic Results on $q_{v,c}(t)$ with Highly Effective Defense

Since the distribution function $q_{v,c}(t)$ of random variable $T_{v,c}$ is hard to characterize in general, we want to know if there are any special cases where we can characterize $q_{v,c}(t)$. In this Section, we present such a result for the special case of highly effective defense, which is abstracted as $c_1, c_2 \to \infty$.

### 3.2.1 Asymptotic results on $q_{v,\mathbf{c}}(t)$ when $c_1, c_2 \to \infty$

Suppose the means of attack inter-arrival times $Y^{(1)}(r)$ and $Y^{(2)}(\theta)$, respectively denoted by $\mu_r$ and $\nu_\theta$, are finite.

PROPOSITION 4. (asymptotic expression of $q_{v,\mathbf{c}}(t)$ when $c_1, c_2 \to \infty$) *Suppose Assumption 2 holds. Suppose $\mu_r$ and $\nu_\theta$ are finite. As $c_1, c_2 \to \infty$, we have*

$$q_{v,\mathbf{c}}(t) \sim \tilde{q}_{v,\mathbf{c}}(t) := \sum_{r=0}^{\deg(v)} \pi_{v,r}(1 - e^{-\bar{F}_r^{(1)}(c_1)t/\mu_r}) + \int_0^\infty \left[1 - e^{-\bar{F}_\theta^{(2)}(c_2)t/\nu_\theta}\right] dH_v(\theta).$$

The asymptotic result $\tilde{q}_{v,\mathbf{c}}$ given in Proposition 4 is useful because it allows to compute $q_{v,\mathbf{c}}(t)$ while demanding much less information (i.e., the means of attack inter-arrival times $\mu_r$ and $\nu_\theta$) than to compute $q_{v,\mathbf{c}}(t)$ according to Eq. (8), which demands information about the distributions of the inter-arrival times between push-based attacks as well as the inter-arrivals times between pull-based attacks. The following numerical example confirms that the asymptotic $\tilde{q}_{v,\mathbf{c}}$ converges to $q_{v,\mathbf{c}}$ as $c_1, c_2 \to \infty$.

### 3.2.2 Example

We continue the above example by using the same distributions of $X^{(1)}(r)$ and $X^{(2)}(\theta)$. However, instead of assuming that the attack inter-arrival times $Y^{(1)}(r)$ and $Y^{(2)}(\theta)$ follow the Gamma distribution, we assume that we only know their means $\mu_r = r^{-1}\Gamma(1+\beta)/\Gamma(\beta)$ and $\nu_\theta = \theta^{-1}\Gamma(1+\lambda)/\Gamma(\lambda)$. Figure 4 plots the asymptotic $\tilde{q}_{v,\mathbf{c}}$ and $q_{v,\mathbf{c}}$ for different $c = c_1 = c_2$. We observe that as $c$ increases, $\tilde{q}_{v,\mathbf{c}}$ converges to $q_{v,\mathbf{c}}$. Actually, $\tilde{q}_{v,\mathbf{c}}$ is already fairly close to $q_{v,\mathbf{c}}$ for $c = 8$ or even $c = 5$. This indicates that we can use the asymptotic $\tilde{q}_{v,\mathbf{c}}$, instead of $q_{v,\mathbf{c}}(t)$, for decision-making purpose when $c_1$ and $c_2$ are large. This is valuable because the former requires much less information to compute than the latter.

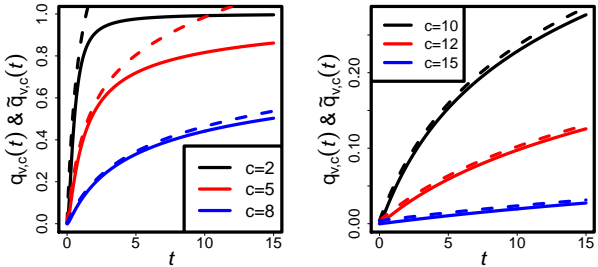

**Figure 4:** $\tilde{q}_{v,\mathbf{c}}(t)$ **(dashed curves) vs** $q_{v,\mathbf{c}}(t)$ **(solid curves):** $c = c_1 = c_2$, $\alpha = 2$, $\lambda = 2.5$, $a = 1$, $b = 2$, $\deg(v) = 8$, **and** $p = .5$. **We observe that the asymptotic** $\tilde{q}_{v,\mathbf{c}}(t)$ **can be used to replace** $q_{v,\mathbf{c}}(t)$ **for** $c \geq 12$ **or even** $c \geq 8$; **this is valuable because the former demands much less information.**

## 3.3 Analyzing Steady-State Compromise Probability $p_{v,\mathbf{c}}$

Now we analyze $p_{v,\mathbf{c}}$, the probability that $v \in V$ is compromised in the steady state. Specifically, let $T_{v,\mathbf{c},i}$ be the length of time interval it takes for $v$ to become the compromised state from the secure state for the $i$th time (i.e., the length of time interval it takes for the $i$th compromise), $R_{v,i}$ be the length of time interval it takes for $v$ to become the secure state from the compromised state for the $i$th time (i.e., the length of time interval it takes for the $i$th recovery), $J_{v,i}$ be the random number of compromised neighbors of node $v$ after the $(i-1)$th recovery, where $i = 1, 2, \ldots$. *Under the condition* that each compromise-and-recovery cycle has the same distribution after each recovery (i.e., reset to the secure state), $J_{v,i}$ for $i = 1, 2, \ldots$ are independently and identically distributed. Therefore, the stochastic process can be seen as an *alternating renewal process* with a sequence of vectors $(T_{v,\mathbf{c},i}, R_{v,i})$ for $i \geq 1$, which have the same distributions as $(T_{v,\mathbf{c}}, R_v)$.

### 3.3.1 Analyzing $p_{v,\mathbf{c}}$ with respect to arbitrary attack-defense graph structure $G$

**How to compute $p_{v,\mathbf{c}}$ numerically?** Recall the attack-defense structure $G = (V, E)$ with $|V| = n$ and adjacency matrix $A = (a_{uv})_{n \times n}$, where $a_{uv} = 1$ if and only if $(u, v) \in E$, and $a_{uv} = 0$ otherwise. Recall also that each $v \in V$ has an associated *alternative renewal process* with i.i.d. time intervals $(T_{v,\mathbf{c},i}, R_{v,i})$, where $T_{v,\mathbf{c},i}$ and $R_{v,i}$ respectively follow distributions $T_{v,\mathbf{c}}$ and $R_v$. The probability that $v$ is compromised in the steady state is [19]:

$$p_{v,\mathbf{c}} = \frac{\mathrm{E}[R_v]}{\mathrm{E}[R_v] + \mathrm{E}[T_{v,\mathbf{c}}]}, \text{ where}$$

$$\mathrm{E}[T_{v,\mathbf{c}}] = \int_0^\infty \mathrm{E}\left[(F_{J_v}^{(1)}(c_1))^{N_{J_v}^{(1)}(t)}\right] \mathrm{E}\left[(F_{\Theta_v}^{(2)}(c_2))^{N_{\Theta_v}^{(2)}(t)}\right] dt.$$

In the steady state, the number of compromised neighbors of $v$ is

$$J_v = \sum_{u=1}^n a_{uv} X_u,$$

where $p_{u,\mathbf{c}} = \mathrm{P}(X_u = 1)$. Thus, we need to solve the following system of equations to get $p_{v,\mathbf{c}}$:

$$p_{v,\mathbf{c}} \int_0^\infty \mathrm{E}\left[(F_{\sum_{u=1}^n a_{uv} X_u}^{(1)}(c_1))^{N_{\sum_{u=1}^n a_{uv} X_u}^{(1)}(t)}\right] \cdot$$
$$\mathrm{E}\left[(F_{\Theta_v}^{(2)}(c_2))^{N_{\Theta_v}^{(2)}(t)}\right] dt - (1 - p_{v,\mathbf{c}})\mathrm{E}[R_v] = 0, \quad v \in V.$$

This system of equations cannot be handled analytically because we are unable to determine the joint distributions of $(X_1, \ldots, X_n)$. Therefore, we resort to the mean-field approximation by replacing random variables $J_v$ and $\Theta_v$ with their respective expectations $\mathrm{E}[J_v]$ and $\mathrm{E}[\Theta_v]$. Since $\mathrm{E}[J_v] = \sum_{u=1}^n a_{uv} p_{u,\mathbf{c}}$, we have

$$\mathrm{E}[T_{v,\mathbf{c}}] = \int_0^\infty \mathrm{E}\left[(F_{\sum_{u=1}^n a_{uv} p_{u,\mathbf{c}}}^{(1)}(c_1))^{N_{\sum_{u=1}^n a_{uv} p_{u,\mathbf{c}}}^{(1)}(t)}\right] \cdot$$
$$\mathrm{E}\left[(F_{\bar{\theta}_v}^{(2)}(c_2))^{N_{\bar{\theta}_v}^{(2)}(t)}\right] dt, \quad (12)$$

where $\bar{\theta}_v = \mathrm{E}[\Theta_v]$. We then obtain the following system of equations with respect to $p_{u,\mathbf{c}}$ for all $v \in V$:

$$p_{v,\mathbf{c}} \int_0^\infty \mathrm{E}\left[(F_{\sum_{u=1}^n a_{uv} p_{u,\mathbf{c}}}^{(1)}(c_1))^{N_{\sum_{u=1}^n a_{uv} p_{u,\mathbf{c}}}^{(1)}(t)}\right] \cdot$$
$$\mathrm{E}\left[(F_{\bar{\theta}_v}^{(2)}(c_2))^{N_{\bar{\theta}_v}^{(2)}(t)}\right] dt - (1 - p_{v,\mathbf{c}})\mathrm{E}[R_v] = 0. \quad (13)$$

By solving the above system of equations for all $v \in V$, we obtain the vector of steady-state compromise probability $(p_{v,\mathbf{c}})_{v \in V}$, which can be used for decision-making purpose.

**How to bound $p_{v,\mathbf{c}}$ under a certain assumption?** Since it is difficult to get closed-form solution of $p_{v,\mathbf{c}}$, we propose to bound $p_{v,\mathbf{c}}$ by using the following assumption.

ASSUMPTION 4. *Suppose the following monotonicity holds:*

(a) $X^{(1)}(r)$ *is increasing in* $r$, *where* $0 \leq r \leq \deg(v)$, *in the usual stochastic sense, meaning that a greater number of* compromised *neighbors implies greater magnitude of push-based attacks.* $X^{(2)}(\theta)$ *is increasing in* $\theta$, *where* $\theta > 0$, *in the stochastic order sense, meaning that severer environment implies greater magnitude of pull-based attacks.*

(b) $Y^{(1)}(r)$ *is decreasing in* $r$, *where* $0 \leq r \leq \deg(v)$, *in the usual stochastic order sense, meaning that a greater number of* compromised *neighbors implies more frequent push-based attacks, and* $Y^{(2)}(\theta)$ *is decreasing in* $\theta$, *where* $\theta > 0$, *in the stochastic sense, meaning that severer environment implies more frequent pull-based attacks.*

PROPOSITION 5. (bounds of $p_{v,\mathbf{c}}$) *Suppose Assumptions 2 and 4 hold. We have for all* $v \in V$

$$\left(1 + \frac{\mathrm{E}[Y^{(2)}(\bar{\theta}_v)]}{\bar{F}^{(2)}_{\bar{\theta}_v}(c_2)\mathrm{E}[R_v]}\right)^{-1} \leq p_{v,\mathbf{c}} \leq \left(1 + (\mathrm{E}[R_v])^{-1} \cdot \int_0^\infty \mathrm{E}\left[(F^{(1)}_{\deg(v)}(c_1))^{N^{(1)}_{\deg(v)}(t)}\right]\mathrm{E}\left[(F^{(2)}_{\bar{\theta}_v}(c_2))^{N^{(2)}_{\bar{\theta}_v}(t)}\right]\mathrm{d}t\right)^{-1}.$$

*Moreover, if* $Y^{(1)}(\deg(v))$ *and* $Y^{(2)}(\bar{\theta}_v)$ *are NBUE for all* $v \in V$, *the following holds for all* $v \in V$:

$$p_{v,\mathbf{c}} \leq \left(1 + \left(\frac{\bar{F}^{(1)}_{\deg(v)}(c_1)\mathrm{E}[R_v]}{\mathrm{E}[Y^{(1)}(\deg(v))]} + \frac{\bar{F}^{(2)}_{\bar{\theta}}(c_2)\mathrm{E}[R_v]}{\mathrm{E}[Y^{(2)}(\bar{\theta})]}\right)^{-1}\right)^{-1}.$$

The bounds given by Proposition 5 are useful because they do not require to solve the system of equations given by Eq. (13).

**How does Proposition 5 (Assumptions 2 and 4) accommodate adaptiveness of attacks?** Note that Assumption 2 is for *fixed* $J_v = r$ and $\Theta_v = \theta$, while Assumption 4 is for *varying* $r$ in the usual stochastic order sense. We here use the usual stochastic order rather than the likelihood ratio order because the latter is a overkill for proving Proposition 5. The adaptiveness accommodated by Proposition 5 is the adaptiveness that is collectively accommodated by Assumptions 2 and 4, namely the adaptiveness obtained by replacing the likelihood ratio order, which was used as an example when we discuss the adaptiveness accommodated by Assumption 1, with the usual stochastic order. More specifically, since $X_i^{(1)}(r)$ is increasing in $r$ for any $i \geq 1$ in the usual stochastic order, for any $s, s' \geq 0$ we have [21]

$$\mathrm{P}(X^{(1)}_{i+1}(J_v) > s'|X^{(1)}_i(J_v) > s) \geq \mathrm{P}(X^{(1)}_{i+1}(J_v) > s'),$$

meaning that a large magnitude $s$ (e.g., a failed sophisticated attack) is more likely followed by another (more) sophisticated attack (when $s' > s$). On the other hand, since $Y_i^{(1)}(r)$ is increasing in $r$, for any $s, s' \geq 0$ we have [21]

$$\mathrm{P}(Y^{(1)}_{i+1}(J_v) \leq s'|Y^{(1)}_i(J_v) \leq s) \geq \mathrm{P}(Y^{(1)}_{i+1}(J_v) \leq s'),$$

meaning that a short attack inter-arrival time (e.g., before launching a failed attack) is more likely followed by another short (or even shorter when $s' < s$) attack inter-arrival time (i.e., the attack can get more intense).

**Remark**. It is worth mentioning that the adaptiveness of attack magnitudes and attack inter-arrival times that is accommodated by Assumption 4 appears within *each* compromise-and-recovery cycle of the renewal process, rather than inter-cycle adaptiveness (which is not accommodated by the model of the present paper — its treatment is left for future research.)

### 3.3.2 Example: $p_{v,\mathbf{c}}$ with respect to regular attack-defense graph structure $G$

As in the examples described above, we set $c = c_1 = c_2$ and assume that the magnitudes of push- and pull-based attacks are Weibull random variables with parameters $(\alpha, 1/r)$ and $(\gamma, 1/\theta)$, and the attack inter-arrival times of push- and pull-based attacks are Gamma random variables respectively with parameters $(\beta, r)$ and $(\lambda, \theta)$, where $\beta, \lambda > 1$, $r$ is the number of $v$'s compromised neighbors, and $\theta$ is the pull-based attack environment.

For regular attack-defense graph structures with $\deg(v) = k$ where $1 \leq k \leq n-1$, we assume that for all $v \in V$, $\bar{\theta}_v = \bar{\theta}$, $\mathrm{E}[R_v] = \mathrm{E}[R]$. Then, $p_{v,\mathbf{c}}$ is the same for all $v \in V$. Let $p_{k,\mathbf{c}}$ be the steady-state compromise probability of any $v$. We have $\sum_{u=1}^n a_{uv}p_{u,\mathbf{c}} = kp_{k,\mathbf{c}}$. Hence, Eq. (13) becomes

$$p_{k,\mathbf{c}} \int_0^\infty \mathrm{E}\left[(F^{(1)}_{kp_{k,\mathbf{c}}}(c_1))^{N^{(1)}_{kp_{k,\mathbf{c}}}(t)}\right] \cdot \mathrm{E}\left[(F^{(2)}_{\bar{\theta}}(c_2))^{N^{(2)}_{\bar{\theta}}(t)}\right]\mathrm{d}t - (1-p_{k,\mathbf{c}})\mathrm{E}[R] = 0,$$

Thus, the lower bound established in Proposition 5 can be expressed as

$$p^-_{k,\mathbf{c}} = \left(1 + \frac{\lambda}{e^{-(\bar{\theta}^{-1}c)^\gamma}\bar{\theta}\mathrm{E}[R]}\right)^{-1}.$$

By noting that $Y^{(1)}(r)$ and $Y^{(2)}(\theta)$ are NBUE, the upper bound is

$$p^+_{k,\mathbf{c}} = \left(1 + \left(\frac{e^{-(k^{-1}c)^\alpha}k\mathrm{E}[R]}{\beta} + \frac{e^{-(\bar{\theta}^{-1}c)^\gamma}\bar{\theta}\mathrm{E}[R]}{\lambda}\right)^{-1}\right)^{-1}.$$

In Table 1, we compute the steady-state compromise probability $p_{k,\mathbf{c}}$ and its bounds $p^-_{k,\mathbf{c}}$, $p^+_{k,\mathbf{c}}$ with different parameters as discussed above. We observe that for fixed $k$, the upper bound $p^+_{k,\mathbf{c}}$ is tighter when $c_1$ and $c_2$ are small (i.e., the networked system is less effectively defended) and the lower bound $p^-_{k,\mathbf{c}}$ is tighter when $c_1$ and $c_2$ are large (i.e., the networked system is more effectively defended). We also observe that for fixed $c_1$ and $c_2$, the upper bound is tighter when $k$ is large.

## 4. RELATED WORK

We classify related prior studies based on their relevance to the goals and means of the present paper. On one hand, a promising approach to modeling and analyzing security of networked systems is centered on modeling the cyber epidemic phenomenon (see, for example, [11, 6, 5, 25, 31, 26, 28]). These studies focused on modeling *push-based* attacks. In order to accommodate drive-by-download attacks, the concept of *pull-based* attacks was studied in [14, 30, 27]. Our goal is to model security of networked systems while offering a unique feature that distinguishes our model from the ones presented in the literature. Specifically, our model can describe the *dynamic* dependence between the relevant random variables, due to the *random environment* technique we use. The issue of accommodating dependence was first addressed in [26] via the Copulas method, which however can accommodate *static* dependence only. It is also worth mentioning that our model can describe attack sophistication and defense capability. This explicitly distinction has been used in [18, 7] for describing attacks against a single assest (e.g., node or computer), and in [29] for describing *active* attack-defense interactions in networked systems.

Our approach to modeling and analyzing security was inspired by the *random environment* and *shock model* techniques in the Theory of Reliability. The random environment technique has been used to describe the external environment that has an impact on the performance of systems and for explaining the dependence between systems that operate in the same environment (see, for example, [23, 16]). (Note that the notion of *random environment* used in the present paper is different from the same term that is however used in a different context [32].) The shock model technique [2, 9, 8, 22, 24, 13, 12] was originally used to describe the phenomenon that systems (or components) may or may not fail under "shocks" of different magnitude. Our model goes beyond the two techniques as used in the Theory of Reliability because of the following. First, we use local random environment to model push-based attacks and global random environment to model pull-based attacks. Second, we allow push-based and pull-based attacks to be against different thresholds, which extend the standard shock model (with only one threshold) and lead to useful analytic results.

| $c$ | $k=5$ | | | $k=8$ | | | $k=10$ | | | $k=12$ | | |
|---|---|---|---|---|---|---|---|---|---|---|---|---|
| | $p_{k,c}$ | $p_{k,c}^-$ | $p_{k,c}^+$ | $p_{k,c}$ | $p_{k,c}^-$ | $p_{k,c}^+$ | $p_{k,c}$ | $p_{k,c}^-$ | $p_{k,c}^+$ | $p_{k,c}$ | $p_{k,c}^-$ | $p_{k,c}^+$ |
| 2.0 | .90 | .87 | .92 | .92 | .87 | .94 | .93 | .87 | .95 | .94 | .87 | .95 |
| 2.5 | .89 | .85 | .91 | .92 | .85 | .93 | .93 | .85 | .94 | .94 | .85 | .95 |
| 3.0 | .88 | .83 | .90 | .91 | .83 | .93 | .93 | .83 | .94 | .94 | .83 | .95 |
| 3.5 | .86 | .82 | .89 | .90 | .81 | .92 | .92 | .82 | .94 | .93 | .82 | .94 |
| 4.0 | .84 | .80 | .87 | .90 | .80 | .92 | .92 | .80 | .93 | .93 | .80 | .94 |
| 5.0 | .79 | .75 | .84 | .88 | .75 | .90 | .91 | .75 | .92 | .92 | .75 | .94 |
| 6.0 | .72 | .70 | .79 | .84 | .70 | .88 | .89 | .70 | .91 | .91 | .70 | .93 |
| 7.0 | .65 | .65 | .73 | .79 | .65 | .86 | .87 | .65 | .90 | .90 | .65 | .92 |
| 8.0 | .59 | .59 | .65 | .67 | .59 | .83 | .83 | .59 | .88 | .89 | .59 | .91 |
| 9.0 | .53 | .53 | .57 | .54 | .53 | .79 | .75 | .52 | .86 | .87 | .53 | .90 |
| $c$ | $k=15$ | | | $k=20$ | | | $k=25$ | | | $k=30$ | | |
| | $p_{k,c}$ | $p_{k,c}^-$ | $p_{k,c}^+$ | $p_{k,c}$ | $p_{k,c}^-$ | $p_{k,c}^+$ | $p_{k,c}$ | $p_{k,c}^-$ | $p_{k,c}^+$ | $p_{k,c}$ | $p_{k,c}^-$ | $p_{k,c}^+$ |
| 2.0 | .95 | .87 | .96 | .96 | .87 | .97 | .97 | .87 | .97 | .97 | .87 | .98 |
| 2.5 | .95 | .85 | .96 | .96 | .85 | .97 | .97 | .85 | .97 | .97 | .85 | .98 |
| 3.0 | .95 | .83 | .96 | .96 | .83 | .96 | .97 | .83 | .97 | .97 | .83 | .97 |
| 3.5 | .95 | .82 | .95 | .96 | .81 | .96 | .97 | .82 | .97 | .97 | .82 | .97 |
| 4.0 | .94 | .80 | .95 | .96 | .80 | .96 | .97 | .80 | .97 | .97 | .80 | .97 |
| 5.0 | .94 | .75 | .95 | .96 | .75 | .96 | .97 | .75 | .97 | .97 | .75 | .97 |
| 6.0 | .94 | .70 | .94 | .95 | .70 | .96 | .96 | .70 | .97 | .97 | .70 | .97 |
| 7.0 | .93 | .65 | .94 | .95 | .65 | .96 | .96 | .65 | .97 | .97 | .65 | .97 |
| 8.0 | .92 | .59 | .93 | .95 | .59 | .95 | .96 | .59 | .96 | .97 | .59 | .97 |
| 9.0 | .92 | .53 | .93 | .95 | .53 | .95 | .96 | .52 | .96 | .97 | .53 | .97 |

Table 1: Steady-state compromise probability $p_{k,c}$ vs. its bounds $p_{k,c}^-$, $p_{k,c}^+$ with parameters $c = c_1 = c_2$, $\alpha = 2$, $\beta = 3.5$, $\gamma = 1$, $\lambda = 1.5$, $\bar{\theta} = 4$, $\mathrm{E}[R_v] = 4$: We observe that for fixed $k$, the upper bound is tighter when $c = c_1 = c_2$ is small and the lower bound is tighter when $c$ is large, and for fixed $c$, the upper bound is tighter when $k$ is large. These indicate the parameter regimes where the lower or upper bound can be used for decision-making purpose.

Finally, it is worthwhile to mention that our goal and approach are different from the goal and approach of *attack graph*, which was initiated by [17, 10, 1] and followed by many others. Attack graph studies assume that the system vulnerabilities are known. In contrast, we do not assume that the vulnerabilities are known (e.g., all known vulnerabilities may have been patched).

## 5. LIMITATIONS OF THE MODEL

First, we represent random environments $J_v$ and $\Theta_v$ as random variables. A more general and powerful representation is that they are driven by stochastic processes. Similarly, we represent the recovery time $R_{v,i}$ as random variable, and it would be better to represent as driven by some stochastic process.

Second, a node can only be compromised by one attack at a time. In practice, a computer can be compromised by multiple attacks during a period of time. While our model can be seen as an approximation to the reality — by considering the aggregate effect of attacks such that a computer is secure only when it is not compromised by any attack, it is interesting to explicitly consider the case that a computer can be compromised by multiple attacks. For this purpose, we may take advantage of the recent model [31], which uses a different approach to accommodate that a node can be compromised by multiple attacks.

Third, we represent the defense thresholds, $c_1$ against push-based attacks and $c_2$ against pull-based attacks, as some deterministic values. A more general and powerful representation is that they are represented by random variables or even driven by some stochastic processes. Moreover, the success of attacks can be probabilistic, as in [29], rather than deterministic.

## 6. CONCLUSION

We have presented a stochastic model for analyzing security of networked systems. The model uses random environment to accommodate a certain degree of adaptiveness of attacks, while using magnitude and threshold to respectively abstract the attack and defense capabilities. The model leads to two natural security metrics, which guide our analysis of the model. We discussed the limitations of the model, which need to be addressed by future research.

**Acknowledgement**. We thank the reviewers for their comments that helped us improve the paper. This work was supported in part by ARO Grant # W911NF-12-1-0286.

# APPENDIX

*Proof of Proposition 1*: Since $\tilde{N}_r^{(1)}(t)$ is a non-homogeneous Poisson process with rate $Q_r(t) = -\log \bar{G}_r^{(1)}(t)$, we have

$$P(\tilde{N}_r^{(1)}(t) \geq m) = \sum_{j=m}^{\infty} \frac{[Q_r(t)]^j}{j!} e^{-Q_r(t)}, \ m \geq 0.$$

Since $Y^{(1)}(r)$ has NBU distribution, from Theorem 3.2 of [3] (p.162) it follows that

$$P(N_r^{(1)}(t) \geq m) \leq P(\tilde{N}_r^{(1)}(t) \geq m)$$

for all $m \geq 0$, which immediately implies that

$$E\left[(F_r^{(1)}(c_1))^{N_r^{(1)}(t)}\right] \geq E\left[(F_r^{(1)}(c_1))^{\tilde{N}_r^{(1)}(t)}\right]$$

because $s^x$ is decreasing in $x > 0$ for $0 < s < 1$. Then,

$$E\left[(F_r^{(1)}(c_1))^{\tilde{N}_r^{(1)}(t)}\right] = e^{-\bar{F}_r^{(1)}(c_1) Q_r(t)} = \left[\bar{G}_r^{(1)}(t)\right]^{\bar{F}_r^{(1)}(c_1)}$$

is the probability generating function of $\tilde{N}_r^{(1)}(t)$ at $F_r^{(1)}(c_1)$. Thus, we have

$$E\left[(F_{J_v}^{(1)}(c_1))^{N_{J_v}^{(1)}(t)}\right] \geq E\left[(\bar{G}_{J_v}^{(1)}(t))^{\bar{F}_{J_v}^{(1)}(c_1)}\right].$$

For the pull-based attacks, we can similarly have

$$E\left[(F_{\Theta_v}^{(2)}(c_2))^{N_{\Theta_v}^{(2)}(t)}\right] \geq E\left[(\bar{G}_{\Theta_v}^{(2)}(t))^{\bar{F}_{\Theta_v}^{(2)}(c_2)}\right].$$

By combing the last two inequalities with Eq. (8), the proposition follows. □

*Proof of Proposition 2*: It is easy to check that both $\left(F_r^{(1)}(c_1)\right)^m$ and $\left(F_\theta^{(2)}(c_2)\right)^m$ in Eqs. (8) are discrete NBUE. By Theorem 2.3 in [4] and the NBUE property of $Y_r^{(1)}$ and $Y_\theta^{(2)}$, we conclude that both $T_{c_1}^{(1)}(r)$ and $T_{c_2}^{(2)}(\theta)$ are NBUE. By using Eq. (4), it can be shown

$$\begin{aligned}
E[T_c(r,\theta)] &= E\left[T_{c_1}^{(1)}(r) \wedge T_{c_2}^{(2)}(\theta)\right] \\
&= \int_0^\infty P\left(T_{c_1}^{(1)}(r) \wedge T_{c_2}^{(2)}(\theta) > t\right) dt \\
&= \int_0^\infty P\left(T_{c_1}^{(1)}(r) > t\right) P\left(T_{c_2}^{(2)}(\theta) > t\right) dt \\
&\geq (E^{-1}[T_{c_1}^{(1)}(r)] + E^{-1}[T_{c_2}^{(2)}(\theta)])^{-1} \\
&= \left(\frac{\bar{F}_r^{(1)}(c_1)}{E[Y^{(1)}(r)]} + \frac{\bar{F}_\theta^{(2)}(c_2)}{E[Y^{(2)}(\theta)]}\right)^{-1}.
\end{aligned}$$

Further, we observe that

$$
\begin{aligned}
&\mathrm{E}[T_{v,\mathbf{c}}] \\
&= \mathrm{E}\left[\mathrm{E}\left[T_{v,\mathbf{c}}|J_v,\Theta_v\right]\right] \\
&\geq \mathrm{E}\left[(\mathrm{E}^{-1}[T_{c_1}^{(1)}(J_v)] + \mathrm{E}^{-1}[T_{c_2}^{(2)}(\Theta_v)])^{-1}|J_v,\Theta_v\right] \\
&= \sum_{r=0}^{\deg(v)} \pi_{v,r} \int_0^\infty \left(\frac{\bar{F}_r^{(1)}(c_1)}{\mathrm{E}[Y^{(1)}(r)]} + \frac{\bar{F}_\theta^{(2)}(c_2)}{\mathrm{E}[Y^{(2)}(\theta)]}\right)^{-1} \mathrm{d}H_v(\theta).
\end{aligned}
$$

The desired result follows immediately. □

*Proof of Proposition 3*: One just note that the Assumption 3 implies that $\prod_{i=1}^m F_{i,r}^{(1)}(c_1)$ and $\prod_{i=1}^m F_{i,\theta}^{(2)}(c_2)$ in Eqs. (5) and (6) are discrete NBUE, with the assumption that the NBUE property of $Y_i^{(1)}(r)$ and $Y_i^{(2)}(\theta)$ as well as $\mathrm{E}[Y_i^{(1)}(r)]$ and $\mathrm{E}[Y_i^{(2)}(\theta)]$ are decreasing in $i \geq 1$, the rest of the proof can be completed very similarly to that of Proposition 2. □

*Proof of Proposition 4*: According to Theorem 1.A4 in [22], it holds that

$$\frac{\bar{F}_r^{(1)}(c_1) T_{c_1}^{(1)}(r)}{\mu_r} \Rightarrow \xi \quad \text{as } c_1 \to \infty,$$

where $\xi$ is a standard exponential random variable, and "$\Rightarrow$" means convergence in distribution. This means that

$$\lim_{c_1 \to \infty} \frac{\mathrm{P}(T_{c_1}^{(1)}(r) \leq \mu_r t / \bar{F}_r^{(1)}(c_1))}{1 - e^{-t}} = 1.$$

The uniformity of convergence in distribution implies

$$\lim_{c_1 \to \infty} \frac{\mathrm{P}(T_{c_1}^{(1)}(r) \leq t)}{1 - e^{-\bar{F}_r^{(1)}(c_1) t / \mu_r}} = 1,$$

and then

$$\lim_{c_1 \to \infty} \frac{\sum_{r=0}^{\deg(v)} \pi_{v,r} \mathrm{P}(T_{c_1}^{(1)}(r) \leq t)}{\sum_{r=0}^{\deg(v)} \pi_{v,r}(1 - e^{-\bar{F}_r^{(1)}(c_1) t / \mu_r})} = 1,$$

i.e.,

$$\mathrm{P}(T_{c_1}^{(1)}(J_v) \leq t) \sim \sum_{r=0}^{\deg(v)} \pi_{v,r}(1 - e^{-\bar{F}_r^{(1)}(c_1) t / \mu_r}). \quad (14)$$

Similarly, for $T_{c_2}^{(2)}(\Theta_v)$, it can be proven to have

$$\mathrm{P}(T_{c_2}^{(2)}(\Theta_v) \leq t) \sim \int_0^\infty \left[1 - e^{-\bar{F}_\theta^{(2)}(c_2) t / \nu_\theta}\right] \mathrm{d}H_v(\theta). \quad (15)$$

Note that from Eq. (4), it follows that for all $c_1, c_2$ and $t > 0$

$$
\begin{aligned}
q_{v,\mathbf{c}}(t) &= \mathrm{P}(T_{c_1}^{(1)}(J_v) \leq t) + \mathrm{P}(T_{c_2}^{(2)}(\Theta_v) \leq t) \\
&\quad - \mathrm{P}(T_{c_1}^{(1)}(J_v) \leq t) \mathrm{P}(T_{c_2}^{(2)}(\Theta_v) \leq t).
\end{aligned}
$$

Thus, combining (15) and (14) leads to the following asymptotically equivalent form of $q_{v,\mathbf{c}}(t)$. □

*Proof of Proposition 5*: Under Assumption 2, for any $0 \leq p_{u,\mathbf{c}} \leq 1$, $\mathrm{E}[T_{v,\mathbf{c}}]$ as defined in Eq. (12) satisfies

$$\mathrm{E}[T_{v,\mathbf{c}}] \leq \int_0^\infty \mathrm{E}\left[(F_{\bar{\theta}_v}^{(2)}(c_2))^{N_{\bar{\theta}}^{(2)}(t)}\right] \mathrm{d}t = \frac{\mathrm{E}[Y^{(2)}(\bar{\theta}_v)]}{\bar{F}_{\bar{\theta}_v}^{(2)}(c_2)},$$

where equality follows Eq. (11). By Assumption 4 and noting that $\mathrm{E}[J_v] \leq \deg(v)$ for all $v$, we have

$$\mathrm{E}\left[(F_{\mathrm{E}[J_v]}^{(1)}(c_1))^{N_{\mathrm{E}[J_v]}^{(1)}(t)}\right] \geq \mathrm{E}\left[(F_{\deg(v)}^{(1)}(c_1))^{N_{\deg(v)}^{(1)}(t)}\right].$$

Applying the two equalities in Eq. (13), we obtain the desired bounds.

If $Y^{(1)}(\deg(v))$ and $Y^{(2)}(\bar{\theta})$ are NBUE for all $v \in V$, from Proposition 2 we have

$$
\begin{aligned}
&\int_0^\infty \mathrm{E}\left[(F_{\deg(v)}^{(1)}(c_1))^{N_{\deg(v)}^{(1)}(t)}\right] \mathrm{E}\left[(F_{\bar{\theta}_v}^{(2)}(c_2))^{N_{\bar{\theta}_v}^{(2)}(t)}\right] \mathrm{d}t \\
&\geq \left(\frac{\bar{F}_{\deg(v)}^{(1)}(c_1)}{\mathrm{E}[Y^{(1)}(\deg(v))]} + \frac{\bar{F}_{\bar{\theta}_v}^{(2)}(c_2)}{\mathrm{E}[Y^{(2)}(\bar{\theta}_v)]}\right)^{-1}.
\end{aligned}
$$

This completes the proof. □